\documentclass[aps,reprint,prl,superscriptaddress,showpacs]{revtex4-1}
\usepackage{graphicx}
\usepackage{amsmath}
\usepackage{hyperref}
\usepackage{xcolor}
\usepackage[all]{hypcap}
\hypersetup{
    colorlinks,
    linkcolor={red!50!black},
    citecolor={blue!50!black},
    urlcolor={blue!50!black}
}
\usepackage{siunitx}

\begin{document}

\title{Photofragmentation Beam Splitters for Matter-Wave Interferometry}

\author{Nadine D\"orre}
\affiliation{University of Vienna, Faculty of Physics, VCQ \& QuNaBioS, Boltzmanngasse 5, A-1090 Vienna, Austria}

\author{Jonas Rodewald}
\affiliation{University of Vienna, Faculty of Physics, VCQ \& QuNaBioS, Boltzmanngasse 5, A-1090 Vienna, Austria}
\author{Philipp Geyer}
\affiliation{University of Vienna, Faculty of Physics, VCQ \& QuNaBioS, Boltzmanngasse 5, A-1090 Vienna, Austria}

\author{Bernd v. Issendorff}
\affiliation{University of Freiburg, Faculty of Physics, Stefan-Meier-Strasse 21, D-79104 Freiburg, Germany}
\author{Philipp Haslinger}
\affiliation{University of Vienna, Faculty of Physics, VCQ \& QuNaBioS, Boltzmanngasse 5, A-1090 Vienna, Austria}
\author{Markus Arndt}
\email[]{markus.arndt@univie.ac.at}
\affiliation{University of Vienna, Faculty of Physics, VCQ \& QuNaBioS, Boltzmanngasse 5, A-1090 Vienna, Austria}

\begin{abstract}
Extending the range of quantum interferometry to a wider class of composite nanoparticles requires new tools to diffract matter waves. Recently, pulsed photoionization light gratings have demonstrated their suitability for high mass matter-wave physics. Here we extend quantum interference experiments to a new class of particles by introducing photofragmentation beam splitters into time-domain matter-wave interferometry. We present data that demonstrate this coherent beam splitting mechanism with clusters of hexafluorobenzene and we show single-photon depletion gratings based both on fragmentation and  ionization for clusters of vanillin.
We propose that photofragmentation gratings can act on a large set of van der Waals clusters and biomolecules which are thermally unstable and often resilient to single-photon ionization. 
\end{abstract}

\maketitle

Recent explorations of matter-wave physics with very massive particles~\cite{Hornberger2012} have been motivated by the rising interest in new tests of the quantum superposition principle~\cite{Nimmrichter2013,Eibenberger2013,Romero-Isart2011,Bateman2013} and quantum sensors. This has triggered the question which scheme might be best adapted to diffract complex nanomatter in a coherent way.
Earlier experiments with absorptive masks of light were based on the possibility to prepare dark states in atoms~\cite{Fray2004,Abfalterer1997a}.
The manipulation of composite particles requires, however, mechanisms which are largely independent of internal particle properties or particular resonances.
Matter-wave interferometry with optical absorption gratings in the time domain (OTIMA) has recently been demonstrated with clusters of anthracene molecules~\cite{Haslinger2013}. This scheme~\cite{Nimmrichter2011b,Reiger2006} is scalable to high masses and has been realized for materials that can be ionized by a single photon~\cite{Marksteiner2009a, Schmid2013} of energy 7.9\,eV. This vacuum ultraviolet (VUV) light can be coherently generated by commercially available fluorine excimer lasers. However, the ionization energy of many organic or biological molecules exceeds 8\,eV and is too high for single-photon ionization gratings.

Here we show that the thermal instability of composite particles, which is often a hindrance in physical chemistry and quantum optics experiments, can be exploited to realize a coherent beam splitter for complex matter. We demonstrate specifically how single-photon absorption in the antinodes of a standing light wave can lead to particle heating and fragmentation and, therefore, to a spatially periodic depletion of the cluster beam. Each light grating acts similarly to a mechanical mask and functions as a diffraction element.
The light pulses trigger the depletion and form together an interferometer in the time domain. What counts is the act of measurement in each grating, which labels a periodic set of particles in the beam, as ``to fragment before detection''. All particles carrying the complementary property ``nonfragmented'' are then read and registered by the detector.

\begin{figure}
\includegraphics[width=\columnwidth]{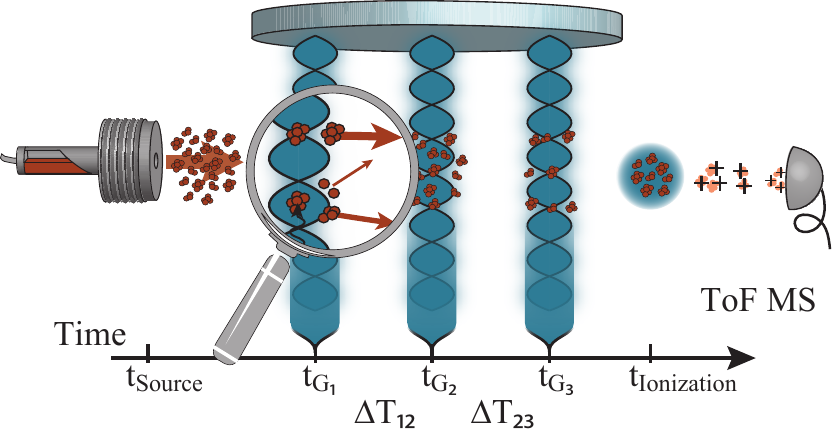}
\caption{Time-domain interference using single-photon fragmentation gratings. 	
A 500\,K pulsed nozzle source emits organic molecules, here hexafluorobenzene (HFB) or vanillin. Supersonic expansion in an intense neon pulse leads to the formation of clusters. Three standing light-wave gratings form the matter-wave interferometer. At the antinodes of the light gratings, the clusters may fragment or ionize after absorption of a single 7.9\,eV (VUV) photon.  This leads to a pulsed and spatially periodic labeling of clusters and their effective removal from the beam. Only clusters transmitted through the absorptive light comb contribute to the interference pattern.}
\label{fig:sketch}
\end{figure}

The experimental setup (see Fig.\,\ref{fig:sketch}) has been described in~\cite{Haslinger2013,Nimmrichter2011b}. Molecules are evaporated and emitted by a pulsed (20\,$\mu$s, 100\,Hz) Even-Lavie valve~\cite{Even2000} to form van der Waals clusters during adiabatic cooling in a coexpanding noble seed gas. The particle cloud passes in close proximity to a two inch dielectric mirror where it is subjected to three VUV laser light pulses (7\,ns, $\lambda_L$  = 157.63\,nm, 3\,mJ in 1$\times$10\,mm$^2$). The light forms standing waves upon retroreflection at the mirror surface. In order to impose spatial matter-wave coherence onto the incident cluster beam, the first grating pulse $G_1$ must be absorptive; i.e., particles in the antinodes must be removed from the detected signal with high efficiency. The node regions then act as sources for elementary matter wavelets. If these sources are sufficiently small, the emerging waves will expand coherently to overlap several nodes and antinodes in the second grating $G_2$. A cluster density pattern forms by virtue of the Talbot-Lau effect as a self-image of $G_2$ which is sampled by the absorptive third grating $G_3$~\cite{Hornberger2012,Nimmrichter2011b}.

While the interference contrast is only determined by the absorptive (depleting) character of the cluster-light interaction in $G_1$ and $G_3$, we also need to consider the dipole interaction between the laser light field and the cluster's optical polarizability in $G_2$. This coupling  imprints a spatially periodic phase onto the matter wave in addition to the amplitude modulation that is caused by depletion~\cite{Haslinger2013}. The particles that are transmitted through the interferometer are ionized by 157.63\,nm light (10\,ns, 0.2\,mJ in $1\times3$\,mm$^2$) and analyzed in a time-of-flight mass spectrometer (ToF-MS). For interference measurements, the power in each grating is adjusted such that less than 25$\%$ of the particles are transmitted. This determines the opening fraction of the grating. The pulse energy of the center grating can be attenuated \textit{in situ} using a 10\,mm long pressure cell which allows us to vary the amount of air in a segment of the evacuated beam line. Since oxygen strongly absorbs in the VUV~\cite{Yoshino2005}, a variation of the air pressure inside the cell between $10^{-4}$ and 200\,mbar is sufficient to reduce the incident laser energy from 90 to almost 0$\%$. In order to monitor pulse-to-pulse variations of the laser power, we use GaP photodiodes to record the relative power of all laser pulses which we cross correlate with the detected ion signal.

\begin{figure}
\includegraphics[width=\columnwidth]{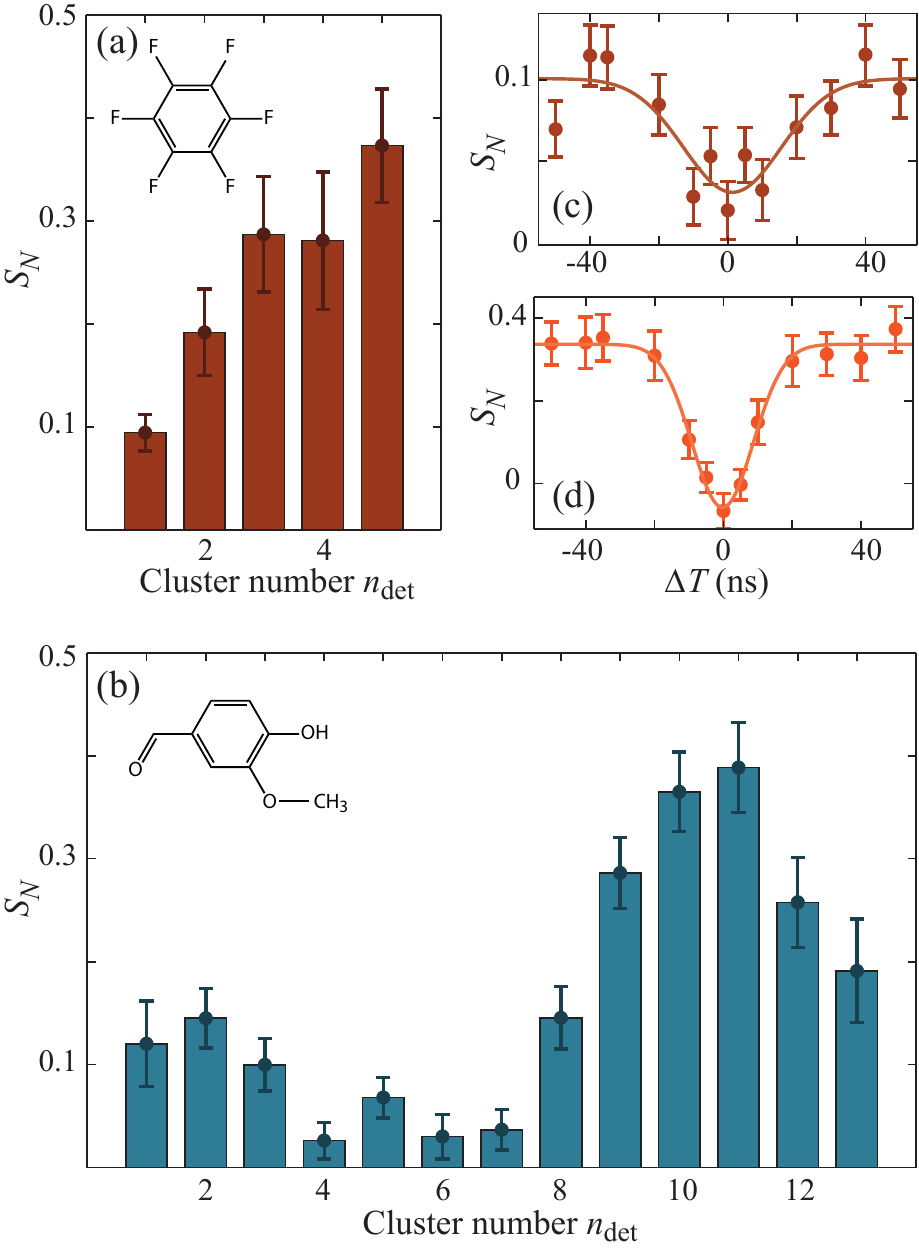}
\caption{Quantum interference of  clusters of hexafluorobenzene (a) and  vanillin (b). We monitor the normalized signal contrast ($S_N$) which compares the cluster transmission for the on resonant and off resonant setting of the grating pulse separation times.  Resonances can be seen in the mass spectrum when the pulse separation time is close to an integer multiple of the Talbot time $T_T$. 	
(c, d)~The resonant character of quantum interference can be seen by varying the difference of the two pulse separation times $\Delta T = \Delta T_{12}-\Delta T_{23}$ between subsequent diffraction gratings in the reference mode. Interference occurs only when both times are equal. A temporal detuning of several dozen nanoseconds suffices to destroy the effect. The dips in (c) and (d) were measured for the detected cluster number $n_{\mathrm{det}}$\,=\,1 and $n_{\mathrm{det}}$\,=\,5 of HFB. The error bars represent one standard deviation of statistical error. The solid lines are Gaussian fits.
}
\label{fig:interference}
\end{figure}

The three grating laser pulses form a time-domain Talbot-Lau interferometer if the delays between two pulses are equal~\cite{Haslinger2013}. This pulse separation time is related to the interfering mass $m$ via the Talbot time $T_T=md^2/h$, where $d=\lambda_L/2$ is the grating period and $h$ is Planck's constant. Matter-wave interference can then be seen in the intensity modulation of the mass spectrum (see Fig.\,\ref{fig:interference})~\cite{Haslinger2013}.
The signal is measured in two complementary modes: an interference mode ($S_{\mathrm{Int}}$) in which the grating pulse separation times are equal, $\Delta T_{12} =\Delta T_{23}$ (Fig.~\ref{fig:sketch}), and a reference mode ($S_{\mathrm{Ref}}$) in which the two times differ by several tens of nanoseconds, $\Delta T_{12}  =\Delta T_{23}+\Delta T$, so that no matter-wave interference can be measured. This is used to express the visibility of the interference pattern in terms of the normalized signal contrast  $S_N= (S_{\mathrm{Int}}-S_{\mathrm{Ref}})/S_{\mathrm{Ref}}$.

We use clusters of hexafluorobenzene ($m$=186\,u per monomer) and vanillin ($m$=152\,u per monomer) as examples for nanoparticles with ionization energies above or close to the grating's photon energy. The vertical ionization energy of hexafluorobenzene (HFB) and vanillin monomers are 9.97\,eV~\cite{Bralsford1960} and 8.30\,eV~\cite{Ponomarev1982}, respectively. Although the ionization potential may fall with increasing cluster size, measurements on benzene indicate that for organic clusters it will not fall by more than 10$\%$ below the value of the monomer~\cite{Krause1991}.

Single-photon ionization is energetically excluded for small HFB and vanillin clusters. Nevertheless, we observe a substantial interference contrast $S_N$ as a function of the detected cluster mass for both species (Fig.\,\ref{fig:interference}). The separation time between the gratings was set to 11.5\,$\mu$s and 18.9\,$\mu$s, respectively, which corresponds to the Talbot time of the fourfold cluster of HFB and the eightfold cluster of vanillin.
For HFB, we have also measured the temporal width of the interference resonance [Figs.\,\hyperref[fig:interference]{2(c)} and \hyperref[fig:interference]{2(d)}]. As expected for time-domain Talbot-Lau interference~\cite{Haslinger2013}, high contrast is only observed if the delay between two grating pulses is equal to within a few nanoseconds. In this case, the interference signal and the reference signal are identical and $S_N$ vanishes.

Demonstrating fragmentation as the cause of the beam splitting process is challenging since the depletion mechanism leaves no trace in the final interference pattern. Ideally, the detector records only those particles that have \emph{not} absorbed a photon in any of the gratings. In addition, the beam splitting angle and the momentum transfer to the particles depend only on the grating geometry and the particle polarizability.
Clusters that absorb a photon may either ionize (followed by extraction from the beam with an external electric field) or fragment. The fragments are unaffected by the field; however, assuming evaporation in thermal equilibrium, the cluster fragments will reach an escape velocity beyond 100\,$\mathrm{m s^{-1}}$. At a forward cluster velocity of 900\,$\mathrm{m s^{-1}}$, the majority of all parent clusters and molecules are therefore ejected beyond the detector acceptance angle of 10\,mrad.

In order to corroborate the beam splitter mechanism, we first show that photoionization requires at least two photons while depletion in the light gratings is a single-photon effect.
\begin{figure}
\includegraphics[width=\columnwidth]{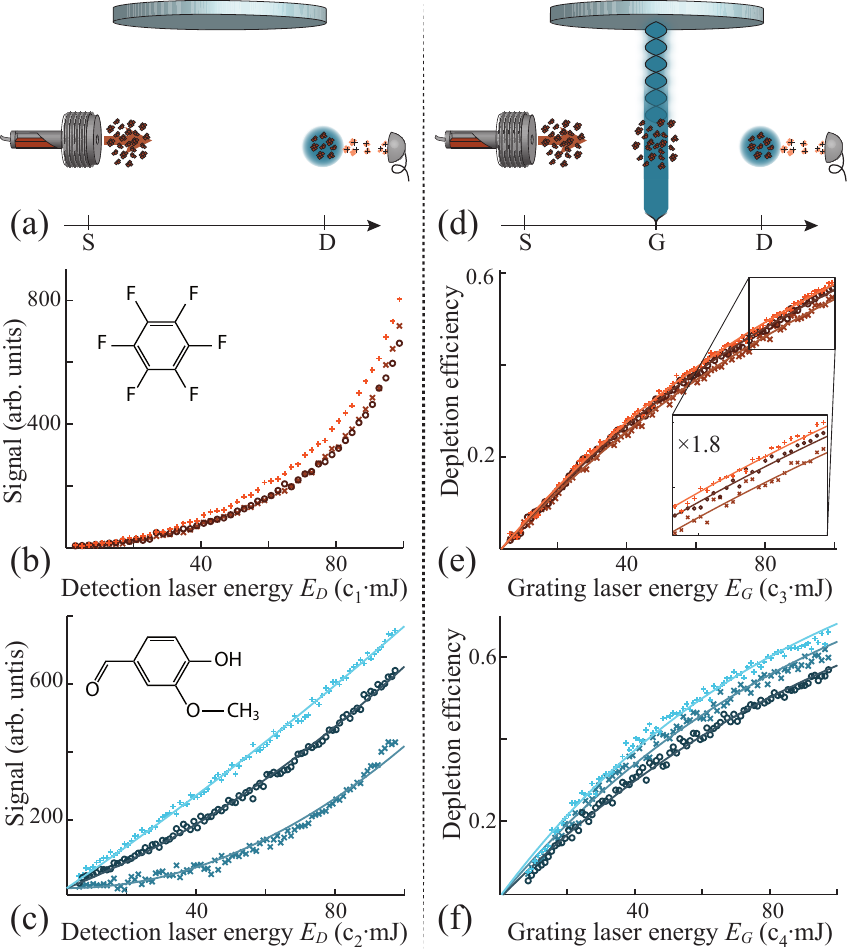}
\caption{
(a) Verifying the multiphoton character of the cluster ionization process: clusters propagate freely between the source 
$S$ and the detector $D$ where they are ionized upon photoabsorption. 
(b)~Photoionization of HFB clusters: the detected ion signal depends nonlinearly on the laser energy $E_D$. This is a clear sign of a multiphoton process. The curves represent
$n_{\mathrm{det}}$\,=\,2\,($\times$, orange), 4\,($\circ$, dark orange), 5\,($+$, light orange)  HFB molecules per detected cluster.
(c)~Photoionization of vanillin clusters: $n_{\mathrm{det}}$\,=\,3\,($\times$, blue), 7\,($\circ$, dark blue), 11\,($+$, light blue). We observe a high-order nonlinearity for $n_{\mathrm{det}}$\,=\,3 and an indication of single-photon events at $n_\mathrm{{det}}$\,=\,11. This is consistent with the expectation that the ionization efficiency increases with increasing cluster number.
(d) Verifying the single-photon character of the beam depletion process in the laser grating $G$: The mirror has been retracted beyond the coherence length of the $F_2$ laser to limit the cluster-light interaction to absorption in a running wave. 
(e) Beam depletion, case of HFB: the curves are well reproduced by single-photon events in a Poissonian process (exponential fits) for the detected cluster numbers $n_{\mathrm{det}}$\,=\,2\,($\times$, orange), 4\,($\circ$, dark orange), 5\,($+$, light orange).  
(f) Beam depletion, case of vanillin: $n_{\mathrm{det}}$\,=\,3\,($\times$, blue), 7\,($\circ$, dark blue), 11\,($+$, light blue). Similarly to HFB, vanillin also shows single-photon fragmentation events.
The constants $c_1$,...,$c_4$ define the scale of the measured laser energy and are different for all four panels.}
\label{fig:transmissiondepletion}
\end{figure}
For that purpose, we have recorded the cluster intensity as a function of the detection laser energy $E_D$ as sketched in Fig.\,\hyperref[fig:transmissiondepletion]{3(a)}. For HFB, we observe a strongly nonlinear power dependence in Fig.\,\hyperref[fig:transmissiondepletion]{3(b)} for all detected clusters at low laser energy consistent with a resonantly enhanced single-photon absorption cross section at 157\,nm~\cite{Motch2006} and a multiphoton ionization process~\cite{Brucat1986}. The detected cluster distribution $S(n_{\mathrm{det}})$  must therefore differ from the incident cluster distribution $S(n_{\mathrm{inc}})$ since fragmentation in the ionization stage depletes larger clusters and replenishes the signal intensity at smaller cluster numbers.

For small clusters of vanillin, we also observe a nonlinear power dependence which gradually changes to a linear one-photon behavior for larger clusters [Fig.\,\hyperref[fig:transmissiondepletion]{3(c)}]. We attribute this transition to the small difference between the photon energy (7.9\,eV) and the ionization energy of the vanillin monomer (8.3\,eV), which will be further reduced for large clusters. In the limit of small laser energy $E_D$, the signal can be expanded to second order: $S_{\mathrm{ion}} \sim A∙E_D+B∙E_D^2$. For small clusters the power series is dominated by the quadratic term ($A$\,=\,0) whereas for n\,$>$\,3,  the emergence of a linear component indicates a one-photon contribution, too.

If the matter-wave beam splitters were dominated by multiphoton processes, we should see a similar nonlinear dependence in the reduction of the cluster transmission as a function of the laser energy $E_G$ in $G_2$.  In order to compare ionization and transmission data, we have reduced the grating to two counter propagating running waves by shifting the interferometer mirror beyond the coherence length of the grating laser [Fig.\,\hyperref[fig:transmissiondepletion]{3(d)}].
The observed beam depletion in Fig.\,\hyperref[fig:transmissiondepletion]{3(e)} (HFB) and Fig.\,\hyperref[fig:transmissiondepletion]{3(f)} (vanillin) is now well represented by exponential curves for all cluster numbers $n_{\mathrm{det}}$. This is expected for a single-photon depletion process with Poissonian statistics. Since ionization was shown to require at least two photons for HFB clusters, the depletion beam splitting must result from single-photon fragmentation. Molecular dynamics simulations of these clusters using \textsc{mmff94}~\cite{Halgren1996} show that a cluster will dissociate within a few picoseconds upon absorption of a single VUV photon and after the conversion of this energy into the vibrational degrees of freedom. A small cluster can even decompose in all its monomeric constituents.
Photofragmentation in combination with ionization in the ToF-MS detector explains the absence of high cluster peaks with large $n_{\mathrm{det}}$ and the absence of clearly discernible Talbot orders in the normalized contrast [Fig.\,\hyperref[fig:interference]{2(a)}]. 
In particular, charged fragments of larger clusters can account for the observed interference signal of the monomer in Fig.\,\hyperref[fig:interference]{2(c)}.  Two-photon ionization of HFB in the gratings may also contribute to genuine monomer interference.

In contrast to HFB, the vanillin cluster data suggest a gradual transition from single-photon fragmentation to a mixture of single-photon ionization and fragmentation when the cluster number increases. This is consistent with the expectation that the cluster ionization energy decreases with the number of constituent molecules.
Since fragmentation of vanillin clusters is less prevalent than for HFB, we can identify the first three Talbot orders in the mass spectrum of vanillin in Fig.\,\hyperref[fig:interference]{2(b)}. They are peaked around $n_{\mathrm{det}}=11\,(m=1672\,\mathrm{u}),5\,(760\,\mathrm{u}),2\,(304\,\mathrm{u})$ as determined by  the pulse separation time. The maxima are shifted to higher masses with respect to the Talbot time because of the dipole force between the cluster polarizability and the laser light field. The fact that high-$n$ vanillin clusters survive the ionization process supports the hypothesis that single-photon ionization competes favorably with photofragmentation at large $n_{\mathrm{det}}$.

In conclusion, we have demonstrated a new class of matter-wave beam splitters which exploit the dissociation of composite objects for the coherent manipulation of particles.
Here, fragmentation is triggered by the absorption of a single photon. Subsequent absorption events may occur but they modify neither the grating transmission function nor the diffraction pattern any further.
 
One might also invoke multiphoton ionization as an alternative to fragmentation for other classes of particles. Indeed, two-photon ionization at 266 - 280 nm can be a valid option for a range of aromatic molecules, including amino acids and polypeptides. However, single-photon processes are favorable to multiphoton schemes since they avoid nondepleting photoabsorption events and, therefore, maximize the interference contrast.
 
Compared to photoionization~\cite{Reiger2006, Haslinger2013} which can be applied to various types of atoms, clusters and molecules, fragmentation can be the dominant labeling process for weakly bound clusters, biomolecules or nanoparticles whose ionization energy exceeds the photon energy of the light grating.	  
Photodepletion has already been successfully used for cluster spectroscopy, using visible~\cite{Haberland1991,Knickelbein1992a} or even infrared wavelengths~\cite{Gough1978,Vernon1981,Gruene2008}.  

One particularly well-suited example of particles susceptible to photofragmentation beam splitters are doped helium nanodroplets.
Such nanodroplets, have been generated in the targeted mass range between 10$^4$ and 10$^9$\,u~\cite{Toennies2001,Stienkemeier2006} and 
have been routinely used as nanocryostats for molecular spectroscopy~\cite{Toennies2004}. At a typical temperature of about 380\,mK the single-atom evaporation rate is low enough~\cite{Brink1990} not to induce any decoherence by particle emission during the 30\,ms coherence time for 
OTIMA interferometry with $10^6$\,u. Moreover, at this temperature all vibrational modes of the dopant are essentially in their ground state and thermal decoherence is  eliminated~\cite{Hackermuller2004}. Photodepletion works exceedingly well in these systems~\cite{Toennies2001} since the heat capacity of helium is low (7.2\,K/atom) and the absorption even of a green photon suffices to evaporate more than 3000 helium atoms. 

Optical fragmentation gratings may also open a new avenue to ion interferometry with composite particles. 
While mechanical diffraction structures have been successfully used for electron diffraction and interference~\cite{Gronniger2005}, they may exhibit local patch potentials or charges. Optical masks can eliminate this problem, as successfully demonstrated with electrons~\cite{Freimund2001}. Dissociation gratings are the most promising option for realizing absorptive gratings for highly charged composite systems. 

Furthermore, photofragmentation gratings are interesting for many biomolecules. Most of them exhibit ionization energies in the range of 8-12\,eV~\cite{Hanley2009} and absorption would often rather induce fragmentation than ionization~\cite{Becker1995}. VUV induced dissociation is frequently used for mass spectroscopy~\cite{Reilly2009,Talbot2005}. A similar mechanism may therefore also serve in realizing absorption gratings for biomolecules.

We acknowledge support by the European Commission (Grant No. 304886), the European Research Council (Grant No. 320694), and the Austrian science funds (Grant No. W1210-3).  We thank Marcel Mayor, Lukas Felix, and Uzi Even for fruitful discussions.


%


\end{document}